# Emotion-Recognition Using Smart Watch Accelerometer Data: Preliminary Findings

**Juan C. Quiroz**
Department of Computing and
Information Systems
Sunway University
Bandar Sunway, Malaysia
juanq@sunway.edu.my

**Min Hooi Yong**
Department of Psychology
Sunway University
Bandar Sunway, Malaysia
mhyong@sunway.edu.my

**Elena Geangu**
Department of Psychology
Lancaster University
Bailrigg, Lancaster LA1 4YW, UK
e.geangu@lancaster.ac.uk

## Abstract
This study investigates the use of accelerometer data from a smart watch to infer an individual's emotional state. We present our preliminary findings on a user study with 50 participants. Participants were primed either with audio-visual (movie clips) or audio (classical music) to elicit emotional responses. Participants then walked while wearing a smart watch on one wrist and a heart rate strap on their chest. Our hypothesis is that the accelerometer signal will exhibit different patterns for participants in response to different emotion priming. We divided the accelerometer data using sliding windows, extracted features from each window, and used the features to train supervised machine learning algorithms to infer an individual's emotion from their walking pattern. Our discussion includes a description of the methodology, data collected, and early results.

## Author Keywords
Emotion-recognition; accelerometer; supervised learning.

## ACM Classification Keywords
I.5.2 Design Methodology: Pattern analysis.

**Movies**

*Happy*
1. 10 Things I Hate About You (1999)
2. When Harry Met Sally (1989)
3. There's Something about Mary (1998)
4. Monty Python (1975)
5. Modern Times (1936)
6. Love Actually (2003)
7. Wall-E (2008)
8. Benny & Joon (1993)

*Sad*
1. Interstellar (2014)
2. Click (2006)
3. Hachi (2009)
4. Shawshank Redemption (1994)
5. Saving Private Ryan (1998)
6. Marley & Me (2008)
7. The Champ (1979)
8. My Girl (1991)

Table 1: Movies used as happiness and sadness stimuli.

**Introduction**

Research in psychology has shown that the way a person walks reflects that person's current mood (or emotional state) [10]. Prior research projects have determined a user's mood or emotion by analyzing the typing behavior on a smartphone [8,16], by tracking smartphone usage [4,6], and by recording and analyzing speech and other smartphone sensors [13].

Most recently, Cui et al. proposed a new method of identifying emotion based on walking patterns [5]. In their work, the accelerometer signals of a smartphone carried by the user were used for supervised learning to classify three emotions: happy, anger, and neutral. However, the work by Cui et al. suffers from a few shortcomings. First, one smartphone was strapped to the user's wrist, and a second smartphone was strapped to the user's ankle, which was an unrealistic scenario. Second, emotion priming was done with a "funny video" and an "infuriating video," with no other details provided in the description of the study. Thus, it is doubtful that the study participants were sufficiently primed on respective emotions and that the necessary data was recorded to suitably train the machine learning algorithms.

In [18], Zhang et al. continued the work by Cui et al. by conducting a study with 123 participants wearing a smart bracelet in their wrist and another smart bracelet on their ankle. Zhang et al. also focused on happy vs anger. They reported accuracies ranging from 60% as high as 91.3% across all subjects using 10-fold cross-validation.

Similar to [5,18], we present preliminary results of an ongoing user study with the goal of inferring the emotion of an individual based on the accelerometer data from a smart watch. In contrast to prior work, we focus on the binary classification of happy vs sad. We used a mixed-design study (Figure 1) to test two types of stimuli for eliciting emotional responses from participants: audio-visual and audio.

**Methodology**

*Participants*
Fifty young adults participated in this study (43 females, $M = 23.18$ years, $SD = 4.87$). All participants were recruited in a university campus (North-West UK) via announcements on notice boards and word of mouth. Each participant was given £7 for participation. None of the participants reported any vision or hearing difficulties and could walk unassisted.

*Materials*
The study included two types of stimuli: (a) audio-visual and b) audio.

AUDIO-VISUAL
For the audio-visual, clips were selected from commercial movies with the potential of being perceived as having emotional meaning (i.e., sadness and happiness) and to elicit emotional responses. The commercial movies were selected from Gross and Levenson [7], Bartollini [2], Schaefer et al. [15], and from five young adults (4 females, $M = 21.50$ years). Another five participants (3 females, $M = 22.80$ years, $SD = 1.30$) were asked to identify each of these clips on terms of the emotion they felt while watching it, and the intensity of the emotion they felt using a 0-to-10 Likert scale (0: hardly, 10 – very much likely). They were also asked if they had watched that movie before. On average, only one participant had seen that movie

**Music**

*Happy*

| Piece | Composer |
|---|---|
| Carmen: Chanson du toreador | Bizet |
| Allegro—A little night music | Mozart |
| Rondo allegro—A little night music | Mozart |
| Blue Danube | Strauss |
| Radetzky march | Strauss |

*Sad*

| Piece | Composer |
|---|---|
| Adagio in sol minor | Albinoni |
| Kol Nidrei | Bruch |
| Solveig's song – Peer Gynt | Grieg |
| Concerto de Aranjuez | Rodrigo |
| Suite for violin & orchestra A minor | Sinding |

Table 2: Musical pieces used as happiness and sadness stimuli.

before. The participants reported that they felt the emotion intended for all clips (100% accuracy) and the intensity ranged between 5.0 to 6.5 for happy and sad clips respectively. See Table 1 for the clips used in our study.

AUDIO

For the audio stimuli, pieces of classical music were chosen known to elicit happy, sad, and emotionally neutral states as reported by [11]. See Tables 2 and 3 for selected clips.

*Procedure*

This was a mixed-design study in which each participant experienced both emotions (within-subject; happy, sad) and assigned to one of the two stimulus types (between-subject: audio-visual, audio). Participants were further divided into three experience conditions: (1) watching the movie clip prior to walking, (2) listening to the music prior to walking, and (3) listening to music while walking. To counterbalance the order of emotion, each participant was randomly assigned to sad-happy and happy-sad. Each participant was tested individually and the task took approximately 20 minutes to complete. All data was collected between 17:00 and 19:00 h to account for peak foot traffic.

Each participant was first greeted by the experimenter at one end of the corridor and proceeded to put on various items. First, the participant had the heart rate sensor (Polar H7) strapped snugly around their chest. The corresponding watch (Polar M400) was strapped onto the experimenter's wrist. The watch was set to "other indoor" sport profile. Second, the participant strapped a smart watch on their left wrist (Samsung Gear 2). For the smart watch, we developed a Tizen application that recorded accelerometer and gyroscope sensor data. The participant rated their current mood state using Positive and Negative Affect Schedule (PANAS) [17] on a 7" tablet. PANAS contains ten adjectives for positive (e.g. joy) and negative feelings (e.g. anxiety) respectively. Scores can range from 10–50, with higher scores representing higher levels of affect.

For the first two conditions, participants placed a pair of headphones to listen or watch the assigned stimuli (e.g. sad music or happy movie). At the end of the stimulus, the participant walked along a corridor to the end and back to the starting point. Participants were reminded not to make any stops in between. The 250m S-shaped corridor was located on the ground floor of one of the university buildings. The experimenter followed the participant at a 100m distance discreetly to observe behavior and to ensure that heart rate monitoring was captured by the watch. Upon their return, the participant rated their mood using the same PANAS scales. Because of the initial mood induction, we always had a neutral condition in between happy and sad conditions to bring them back to a normal calm state. The same procedure above - rating their initial mood using PANAS, watched or listened to a stimulus, walked along the corridor and back, and rated their mood – was applied to the neutral and second emotion.

In the latter condition – listen while walking – the procedure was similar to the above except that the participant was listening to the assigned music while walking.

**Music**

*Neutral*

| Piece | Composer |
|---|---|
| L'oiseau prophete | Schumann |
| Claire de lune | Beethoven |
| Claire de lune | Debussy |
| Symphony no. 2 C minor | Mahler |
| La traviata | Verdi |
| Pictures at an exhibition | Mussorgsky |
| Water music— passepied | Handel |
| Violin romance no. 2 F major | Beethoven |
| Water music— minuet | Handel |
| The planets— Venus | Holst |

Table 3: Neutral musical pieces.

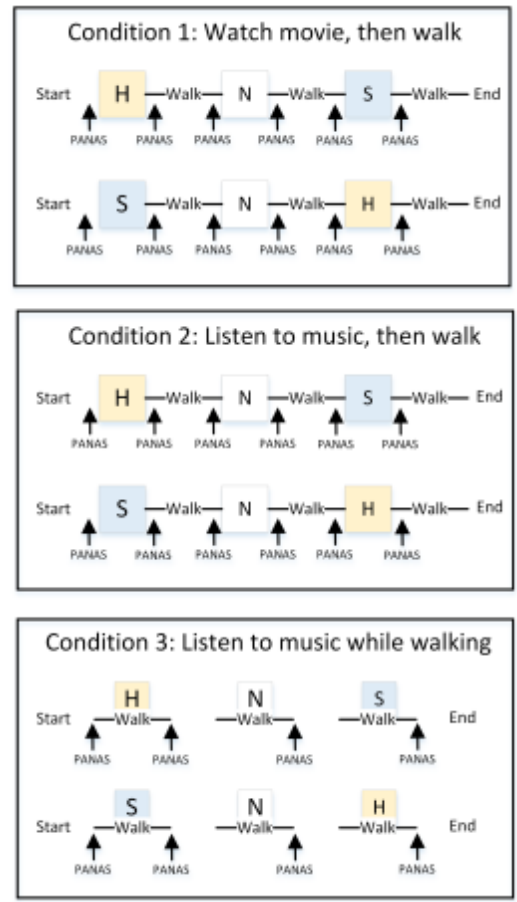

**Figure 1**: Mixed-design study with three conditions.

### Feature Extraction

Participants wore a Samsung Gear S2 smart watch on their wrist for the entire duration of the experiment. The smart watch included a tri-axial accelerometer and a tri-axial gyroscope. The sampling rate of the smart watch is advertised as 25 Hz, but our results show that actual sampling rate was average of 23.8 Hz. We recorded both accelerometer and gyroscope data, but we only present preliminary analysis of accelerometer data in this paper.

During the experiment, the experimenter recorded the time when each participant started and stopped walking. These times where used to identify the accelerometer data that corresponded to actual walking time. Figure 2 illustrates the raw accelerometer data for one participant under condition 2, listen to music and then walk, and stimulus order of happy-neutral-sad. Using the stimulus order, we assigned labels to the features extracted from each of the three walking time periods.

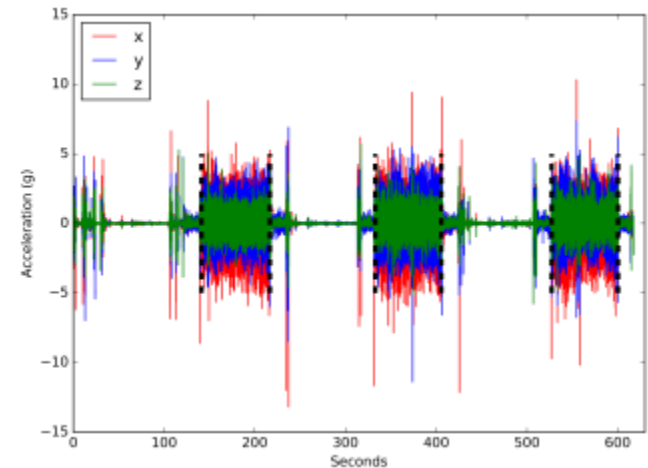

**Figure 2**: Accelerometer data for one participant, condition 2, listen to music and then walk, with the three regions delineated by black dotted lines corresponding to happy-neutral-sad.

**Features Extracted**
1. Mean
2. Max
3. Min
4. Standard deviation
5. Energy
6. Kurtosis
7. Skewness
8. RMS
9. RSS
10. Area under the curve
11. Absolute area under the curve
12. Absolute value mean
13. Range
14. Quartiles
15. Mean absolute deviation
16. Angle between vectors
17. Magnitude

These features have been used in prior wok for activity recognition [1,7].

The start-end walking times were labeled according to the corresponding emotion priming done before walking. For example, if the participant viewed a movie clip for evoking happiness, all of the windows extracted from the subsequent start-end time were labeled as happy. These labels were used to train and test supervised learning algorithms for the binary classification of happiness vs sadness. Because of the initial mood induction, we always had a neutral condition in between happy and sad conditions to bring the participants back to a normal calm state. Thus, our initial motivation was to classify happy vs sad. The classification of happy vs neutral vs sad will be completed in future work.

We first filtered the raw accelerometer data with a mean filter (window=3). Features were extracted from sliding windows without overlap and a window size of one second. Our feature extraction approach is similar to that used for activity recognition from smartphone accelerometer data [1,14]. That is, each window is treated as an independent sample (feature vector).

We divided the data by condition, and built personal models with the features extracted from each window [9]. In personal models, the training and testing data comes from a single user. In our case, we built 44 personal models (the data from 6 participants was discarded because of missing data and other recording errors), with each model evaluated using five-fold cross-validation. For each participant we had an average of 205.30 ± 22.82 samples labeled as happy, 205.55 ± 20.14 samples labeled as sad, and a total of 410.84 ± 38.16 samples per participant. Out of the 44 personal models built, 16 were from condition 1 (watch movie and then walk), 14 were from condition 2 (listen to music and then walk), and 14 were from condition 3 (listen to music and then walk).

We compared random forest, logistic regression, and a baseline classifier that picked the majority class as the prediction. The python scikit-learn library was used for the training and testing of these classifiers. Since the number of samples labeled as happy vs sad for each participant was approximately the same, the baseline classifier predicted each window as happy vs sad with about a 50% probability (i.e. all samples were classified as happy, resulting in about a 50% accuracy). The personal models we built are naïve in that each window is an independent sample. Therefore, a model could potentially predict happy-sad-happy for three consecutive one second windows, which is unrealistic as a user is not likely to go from happy to sad and back to happy in a matter of 3 seconds. This limitation of our modeling approach will be addressed in future work.

**Preliminary Results**
*Behavioural observations*
When asked about their experience in using a smart gadget, most of the participants were familiar and comfortable with the smart watch and smartphone, but not the heart rate monitor. They did not notice anything unusual about the study, which may have influenced their walking gait

*Emotional Response to Stimuli*
We have analyzed the PANAS responses for conditions 1 and 2 for current participants on the happy versus sad stimuli. The neutral state was not used. We leave the analysis of condition 3 for future work.

CONDITION 1: WATCH MOVIE AND THEN WALK
Participants reported a reduced negative affect after watching a sad movie clip (M = 14.94, SD = 6.79) compared to before (M = 19.00, SD = 7.20), t (16) = 3.16, p = .006. There was no significant difference for positive affect for the sad movie, t (16) = .08, p = .94 and for both affect in the other two emotions (happiness and neutral), all ps > .10.

CONDITION 2: LISTEN TO MUSIC AND THEN WALK
For the sad music, participants reported an increased positive affect after the walk (M = 24.00, SD = 5.33) compared to prior (M = 20.31, SD = 5.79), t (15) = 2.96, p = .01, and a reduced negative affect after (M = 11.69, SD = 3.34) as to before (M = 13.63, SD = 5.12), t (15) = 2.78, p = .014. Unlike the sad music, participants reported a reduced positive affect after listening to happy music (M = 26.38, SD = 6.96) compared to before (M = 29.56, SD = 5.17), t (15) = 2.62, p = .02, but no significant difference for the negative affect, t (15) = 1.60, p = .13. There is no significant difference for the neutral music for both affect, both ps > .76.

DISCUSSION
Participants reported feeling less negative affect after watching or listening to the sad stimuli, and developed more positive affect as observed in the music condition but not in the movie condition. These emotional states indicate that the walking is useful to alleviate negative mood, similar to [3,12]. A sub-set of 10 participants reported liking the sad stimulus the least compared to happy and neutral stimuli. This personal preference self-report further adds credence to the PANAS results in that walking is useful in alleviating negative mood. However, we did not examine whether participants' preference for sad stimuli may have influenced their overall emotional state.

*Accelerometer Data*
Figure 3 illustrates boxplots showing the distribution of classification accuracies for each participant separated by condition. For all three conditions, the baseline is usually around 50% accuracy.

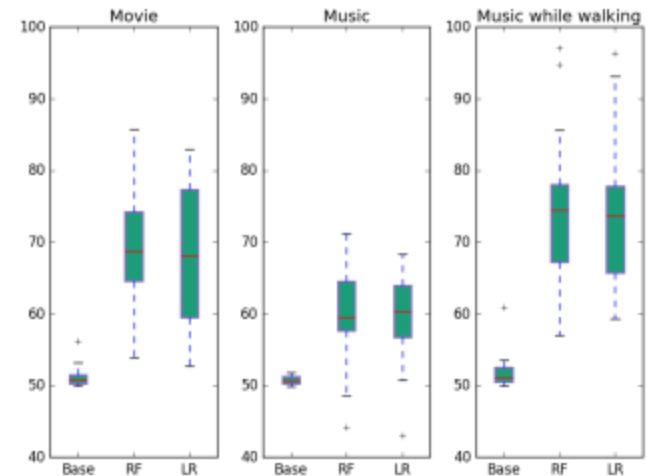

**Figure 3**: Boxplot of classification accuracies for participants divided by conditions. Algorithms tested were baseline (pick majority), random forest, and logistic regression. Outliers are indicated by +.

For all conditions, both random forest and logistic regression outperform the baseline, with accuracies for some users being as low as 50, but with the majority of user accuracies being in the range of 60 to 80. However, we notice that the distributions for the movie stimulus and listening to music while walking results in walking patterns that can be more easily distinguished

(happy vs sad) as opposed to listening to music and then walking. Statistical tests are yet to be conducted.

There are a number of factors that can influence the classification results. These include window size, filtering methods, and choice of features. Our analysis on the basis of all these factors is ongoing.

**Conclusions and Future Work**
We have presented preliminary findings of our ongoing user study on determining the emotions of individuals based on the accelerometer data from a smart watch. Additional participants are being recruited, as bigger sample sizes are needed for drawing appropriate conclusions. We are also in the process of analyzing the heart rate data, the effect of window size, and feature engineering.

Emotional information appears in all forms; face, voice, body posture. To a great extent, our emotions are reflected in our body postures and gait. For example, we are more likely to lean forward and to direct our clenched fists towards our source of frustration when we are angry. An increasing body of evidence emerged in recent years suggest that not only we tend to adopt different body postures and gait according to the emotions we feel, these are easily identified by others as well, reinforcing the idea that they are important social signals. These are important not only for social communication purposes, but also with the scope of monitoring and intervening in cases of non-adaptive emotion reactivity and regulation, particularly in specific clinical conditions. With the increased availability of wearable body sensors that can readily measure and record walking gait, speed and other psychophysiological indices of emotional arousal, it will be interesting to design and test the feasibility of accurately recording steps with different emotional states. Further, other biosensors could be used jointly with measures of skin response and subjective ratings of affective states in biofeedback style interventions for emotion regulation.

Other biosensors could be used jointly with measures of skin response and subjective ratings of affective states in biofeedback style interventions for emotion regulation.

**Acknowledgements**
We thank all the volunteers who participated in our user study. Elisa Roberti and Magdalene Rose for assisting in the data collection.

**References**
1. Anguita. 2013. A Public Domain Dataset for Human Activity Recognition Using Smartphones. .
2. Ellen Elizabeth Bartolini. 2011. Eliciting emotion with film: development of a stimulus set. .
3. Emily E. Bernstein and Richard J. Mcnally. 2016. Acute aerobic exercise helps overcome emotion regulation deficits. *Cognition and Emotion*: 1–10.
4. G. Chittaranjan, J. Blom, and D. Gatica-Perez. 2011. Who's Who with Big-Five: Analyzing and Classifying Personality Traits with Smartphones. *2011 15th Annual International Symposium on Wearable Computers (ISWC)*, 29–36.
5. Liqing Cui, Shun Li, and Tingshao Zhu. 2016. Emotion Detection from Natural Walking. *Human Centered Computing*, Springer, Cham, 23–33.

6. G. Drake, E. Csipke, and T. Wykes. 2013. Assessing your mood online: acceptability and use of Moodscope. *Psychological Medicine* 43, 7: 1455–1464.
7. James J. Gross and Robert W. Levenson. 1995. Emotion elicitation using films. *Cognition and Emotion* 9, 1: 87–108.
8. Hosub Lee, Young Sang Choi, Sunjae Lee, and I.P. Park. 2012. Towards unobtrusive emotion recognition for affective social communication. *2012 IEEE Consumer Communications and Networking Conference (CCNC)*, 260–264.
9. Jeffrey W. Lockhart and Gary M. Weiss. 2014. Limitations with Activity Recognition Methodology & Data Sets. *Proceedings of the 2014 ACM International Joint Conference on Pervasive and Ubiquitous Computing: Adjunct Publication*, ACM, 747–756.
10. Johannes Michalak, Katharina Rohde, and Nikolaus F. Troje. 2015. How we walk affects what we remember: Gait modifications through biofeedback change negative affective memory bias. *Journal of Behavior Therapy and Experimental Psychiatry* 46: 121–125.
11. Martina T. Mitterschiffthaler, Cynthia H.Y. Fu, Jeffrey A. Dalton, Christopher M. Andrew, and Steven C.R. Williams. 2007. A functional MRI study of happy and sad affective states induced by classical music. *Human Brain Mapping* 28, 11: 1150–1162.
12. Frank J Penedo and Jason R Dahn. 2005. Exercise and well-being: a review of mental and physical health benefits associated with physical activity. *Current Opinion in Psychiatry* 18, 2.
13. Kiran K. Rachuri, Mirco Musolesi, Cecilia Mascolo, Peter J. Rentfrow, Chris Longworth, and Andrius Aucinas. 2010. EmotionSense: A Mobile Phones Based Adaptive Platform for Experimental Social Psychology Research. *Proceedings of the 12th ACM International Conference on Ubiquitous Computing*, ACM, 281–290.
14. Jorge-L. Reyes-Ortiz, Luca Oneto, Albert Samà, Xavier Parra, and Davide Anguita. 2016. Transition-Aware Human Activity Recognition Using Smartphones. *Neurocomputing* 171: 754–767.
15. Alexandre Schaefer, Frédéric Nils, Xavier Sanchez, and Pierre Philippot. 2010. Assessing the effectiveness of a large database of emotion-eliciting films: A new tool for emotion researchers. *Cognition and Emotion* 24, 7: 1153–1172.
16. S. Shapsough, A. Hesham, Y. Elkhorazaty, I. A. Zualkernan, and F. Aloul. 2016. Emotion recognition using mobile phones. *2016 IEEE 18th International Conference on e-Health Networking, Applications and Services (Healthcom)*, 1–6.
17. David Watson, Lee A. Clark, and Auke Tellegen. 1988. Development and validation of brief measures of positive and negative affect: The PANAS scales. *Journal of Personality and Social Psychology* 54, 6: 1063–1070.
18. Zhan Zhang, Yufei Song, Liqing Cui, Xiaoqian Liu, and Tingshao Zhu. 2016. Emotion recognition based on customized smart bracelet with built-in accelerometer. *PeerJ* 4: e2258.